\documentclass{svproc}
\usepackage{url}

\usepackage[cmex10]{amsmath}
\usepackage{graphicx}
\usepackage{subfigure}
\usepackage{cite}
\graphicspath{{./Figures/}}
\urldef{\mailsa}\path|danilo.comminiello@uniroma1.it|
\begin{document}
\mainmatter
\title{A Multimodal Deep Network for the Reconstruction of T2W MR Images}
\titlerunning{Multimodal Deep Network}
\author{Antonio Falvo \and Danilo Comminiello \and Simone Scardapane \and \\ Michele Scarpiniti \and Aurelio~Uncini}
\authorrunning{Falvo et al.}
%
%
\institute{Dept. Information Eng., Electronics and Telecommunications (DIET)\\
Sapienza University of Rome, Via Eudossiana 18, 00184 Rome, Italy\\
\mailsa}
\maketitle
%
\begin{abstract}
Multiple sclerosis is one of the most common chronic neurological diseases affecting the central nervous system. Lesions produced by the MS can be observed through two modalities of magnetic resonance (MR), known as T2W and FLAIR sequences, both providing useful information for formulating a diagnosis. However, long acquisition time makes the acquired MR image vulnerable to motion artifacts. This leads to the need of accelerating the execution of the MR analysis. In this paper, we present a deep learning method that is able to reconstruct subsampled MR images obtained by reducing the $k$-space data, while maintaining a high image quality that can be used to observe brain lesions. The proposed method exploits the multimodal approach of neural networks and it also focuses on the data acquisition and processing stages to reduce execution time of the MR analysis. Results prove the effectiveness of the proposed method in reconstructing subsampled MR images while saving execution time.
\end{abstract}
%
%
%
%
%
\section{Introduction}
\label{sec:intro}
Nuclear magnetic resonance (NMR) is a transmission analysis technique that allows to obtain information on the state of matter, exploiting the interaction between magnetic fields and atoms nuclei. In the biomedical field, information deriving from the NMR is represented in the form of tomographic images. Nowadays, the NMR plays an important role in the health field, and it allows to carry out a whole typology of diagnostic exams, from traditional to functional neuroradiology, from internal diagnostic to obstetrics and pediatric diagnostics \cite{Bell1984}. 

During the acquisition stage of an MR signal, it is necessary to sample the entire $k$-space to obtain images that are as much detailed as possible \cite{Haacke1999, Liang2000}. Data in the $k$-space encode information on spatial frequencies and are generally captured line by line. Therefore, the acquisition time for a given sequence depends on the number of lines sampled in the $k$-space, thus leading to a rather slow acquisition process. Moreover, significant artifacts may occur in the MR images caused by to slow movements of the patient, due to physiological factors 
or to fatigue, e.g., too much time in the same position \cite{Haacke1999, Liang2000}. Moreover, the long scan time also increases the healthcare cost for the patient, besides limiting the availability of MR scanners.

Over the years, several methods, such as compressed magnetic resonance and parallel magnetic resonance \cite{LustigMRM2007, GamperMRM2008, LustigSPM2008, JaspanBJR2015}, have been proposed to accelerate MRI scans by skipping the $k$-space phase coding lines and avoid the aliasing phenomenon introduced by subsampling. The problem of accelerating magnetic resonance can also be tackled through deep learning techniques. In particular, the reconstruction of tomographic images has been often efficiently addressed by using convolutional neural networks (CNNs) \cite{JinTIP2017, MccannSPM2017, SchlmperTMI2018, RoyARXIV2018, QinTMI2019}.

Most of the state-of-the-art methods focus on the reconstruction of MR images using a unimodal neural architecture where a subsampled image to be reconstructed is provided as input. In this paper, we propose a new deep learning method for reconstructing MR images by exploiting additional information provided by FLAIR images. Such images are widely used in the MR diagnosis as they allow to enhance the brain lesions due to the disease. FLAIR images are highly correlated with the T2 weighted images (T2WIs), thus the joint use of such images increases the efficiency of the reconstruction and also presents much more information in the lesion region. To exploit both the images, we propose a multimodal deep neural network, inspired by the well-known U-Net, a convolutional-based model that was developed for biomedical image segmentation \cite{RonnenbergerMICCAI2015}. In the literature, several studies have been proposed using a multimodal approach for image reconstruction. In \cite{XiangTBE2018}, T2WIs were attempt to be estimated from T1WI, while other works focus on improving the quality of the subsampled images with the help of high resolution images with different contrast \cite{HuangMRI2014, KimMP2018}. However, to the best of our knowledge, no attempt has ever been made to reconstruct T2WIs from subsampled T2WIs (T2WIsub) and FLAIR images while maintaining high image quality in the area of lesions.

Experimental results prove that the proposed method is able to accelerate the MR analysis four times, while preserving the image quality, with a high detail of any lesion and negligible aliasing artifacts.

The rest of the paper is organized as follows. In Section~\ref{sec:mask}, we introduce the proposed approach, including a new subsampling mask, while the proposed Multimodal Dense U-Net is presented in Section~\ref{sec:multiunet}. Results are shown in Section~\ref{sec:results} and, finally, our conclusion is drawn in Section~\ref{sec:concl}.
%
%
%
%
%
\section{Proposed Approach: Main Definitions}
\label{sec:mask}
We first focus on the images to be provided as input in order to reconstruct the T2WI.
\subsection{Problem Formulation}
\label{subs:optimiz}
We denote with ${\cal{X}}_{T2}$ the $k$-space for the T2WI that represents the target. Multiplying the $k$-space ${\cal{X}}_{T2}$ for a suitably designed mask $M$, it is possible to obtain a subsampled version of the $k$-space, i.e.,
\begin{equation}
	{\cal{X}}_{T2sub} = M \cdot {\cal{X}}_{T2}
	\label{eq:subskspace}
\end{equation}

The bidimensional inverse Fourier transform allows to achieve data into the space domain. Therefore, we define the fully-sampled target image $Y_{T2}$ and the subsampled T2 image $Y_{T2sub}$ to be used for reconstruction through the proposed deep network. Finally, we denote the FLAIR image to be provided as input with $Y_F$.

We want to reconstruct the fully-sampled T2 image $Y_{T2}$, given the only availability of subsampled $Y_{T2sub}$ and $Y_F$. The reconstructed T2 image is denoted as $\hat{Y}_{T2}$. To this end, we build and train a deep network to minimize the following loss function:
\begin{equation}
	\arg \min \left\{\text{MSE} + \text{DSSIM}\right\}
	\label{eq:loss}
\end{equation}

\noindent in which the MSE denotes the mean-square error and DSSIM the structural dissimilarity index. The former is defined as:
\begin{equation}
	\text{MSE} = \frac{1}{N}\sum_{i=1}^{N}\left(Y_{T2,i} - \hat{Y}_{T2,i}\right)^2.
	\label{eq:mse}
\end{equation}

On the other hand, the DSSIM is complementary to the structural similarity index (SSIM), which is often adopted to assess the perceived quality of television and film images as well as other types of digital images and videos. 
It was designed to improve traditional methods such as the signal-to-peak noise ratio (PSNR) and the mean square error (MSE), and it is defined as:
\begin{equation}
	\text{DSSIM}\left(Y_{T2},\hat{Y}_{T2}\right)  = \frac{1}{2} - \frac{\left(2\mu_{Y}\mu_{\hat{Y}} + c_1\right)\left(2\sigma_{Y\hat{Y}} + c_2\right)}{2\left(\mu_{Y}^2 + \mu_{\hat{Y}}^2 + c_1\right)\left(\sigma_{Y}^2 + \sigma_{\hat{Y}}^2 + c_2\right)}
	\label{eq:dssim}
\end{equation}

\noindent where $\mu_{Y}$, $\mu_{\hat{Y}}$ represent the mean values, $\sigma_{Y}^2$ and $\sigma_{\hat{Y}}^2$ the variances, and $\sigma_{Y\hat{Y}}$ the covariance.
\subsection{Customization of a New Subsampling Mask}
\label{subs:submask}
Most of the existing literature dealing with MRI acceleration mainly focuses on the reconstruction of images. However, the quality of the reconstruction depends significantly on how $k$-space is sampled. This problem can be faced essentially by adopting one of the following approaches: 1) a \textit{dynamic approach} based on deep learning 
in which cells of fixed width are allowed to 
move in the $k$-space and change position based on the reconstruction performance;  2) a \textit{static approach} in which fixed sampling masks are used that go to select only certain areas of the $k$-space.

In this work, we choose the static approach since the dynamic one does not guarantee that the power spectrum of a reference image is similar to that of the test image. The adopted sampling method consists of a mask that acts along the direction of the phase coding of the $k$-space, in which it is possible, once the subsampling factor is set, to choose the percentage of samples that will occupy the central part of $k$-space, thus leaving the rest of the samples equidistant from each other.

\begin{figure}[t]
	\centering
	\subfigure[]{\includegraphics[width=0.48\textwidth,keepaspectratio]{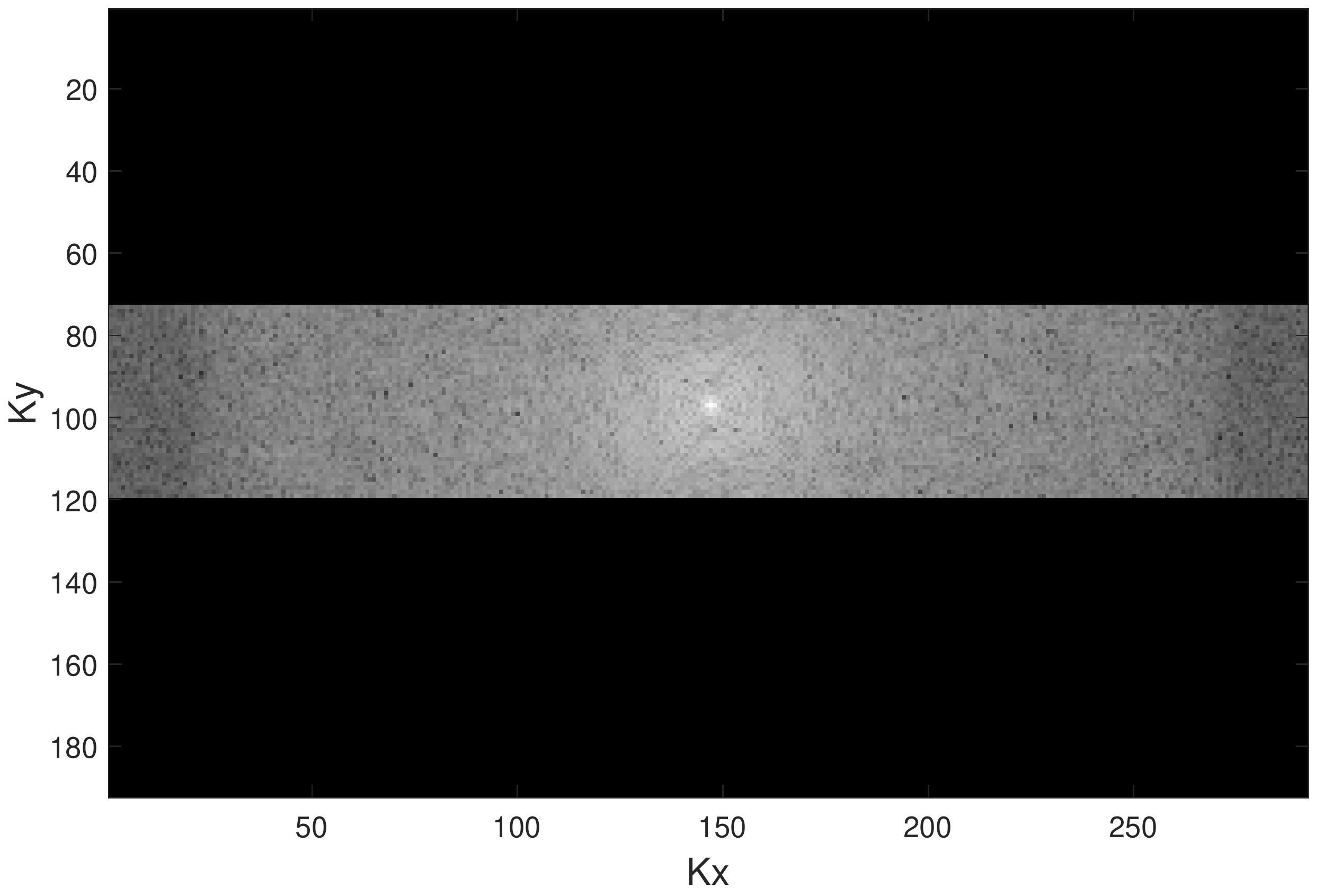}\label{fig:centermask}}
\subfigure[]{\includegraphics[width=0.48\textwidth,keepaspectratio]{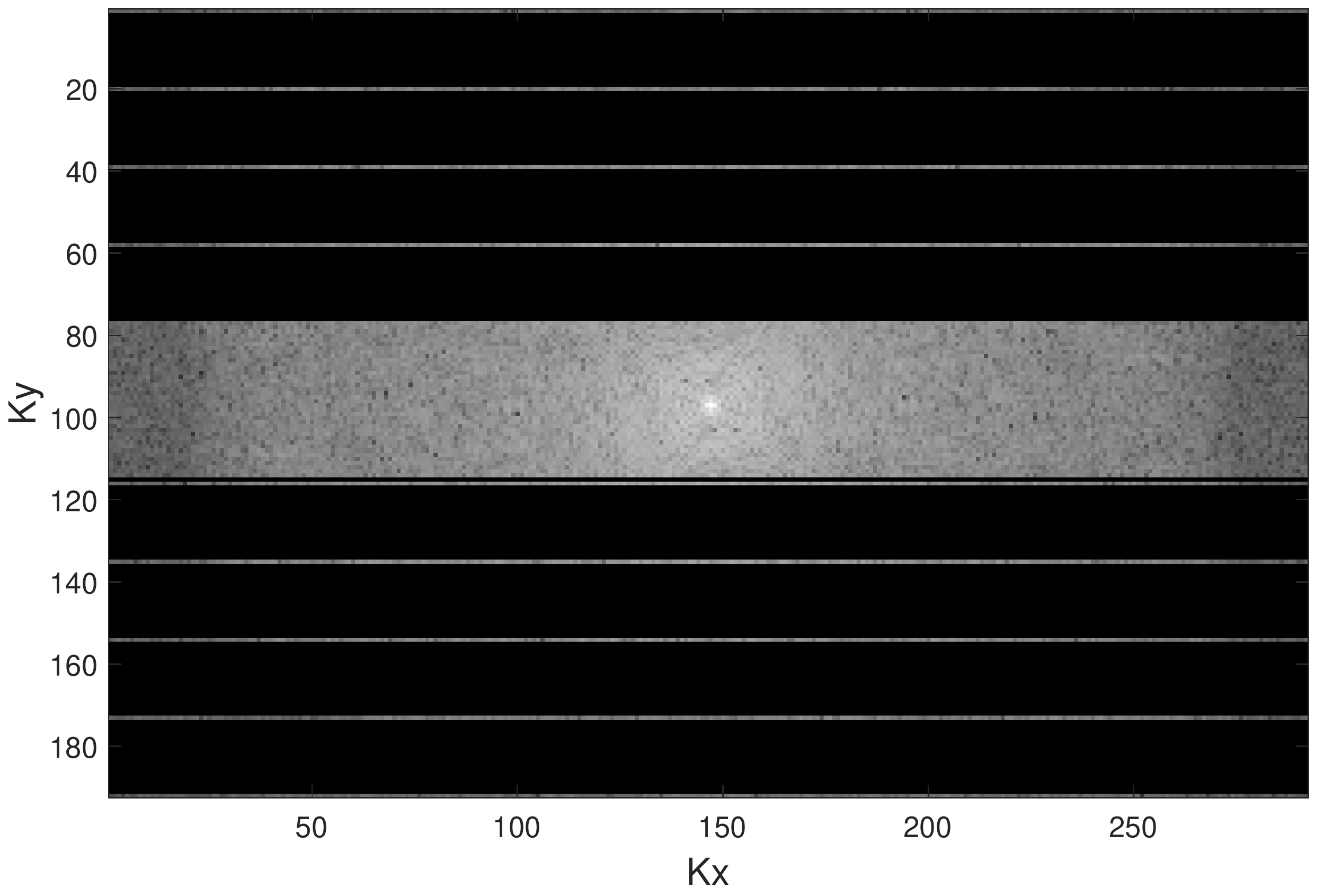}\label{fig:custommask}}
	\caption{Subsampling masks: a) center mask and b) the proposed custom mask.}
	\label{fig:masks}
\end{figure}

Figure~\ref{fig:masks} shows two different types of mask both obtained by setting a subsampling factor $k = 4$. In the center mask of Fig.~\ref{fig:centermask}, samples are taken exclusively in the central area of the $k$-space, where most of the low-frequency components can be found providing useful information on the contrast of the image \cite{XiangTBE2018}. However, in this work, we propose a new mask, depicted in Fig.~\ref{fig:custommask}, which selects the 80\% of the total samples from the center and the remainder in an equidistant manner so as to have information even in the high frequencies.

%
%
%
%
%
%
\section{Multimodal Dense U-Net}
\label{sec:multiunet}
The proposed neural network is multimodal architecture: on one branch we provide the T2WIsub as input while on the other branch we provide the FLAIR image to be used to improve the reconstruction quality. We expect all the spatial information in the FLAIR image to help estimate the anatomical structures in T2WI.

\begin{figure}[t]
	\centering
	\includegraphics[width=0.98\textwidth,keepaspectratio]{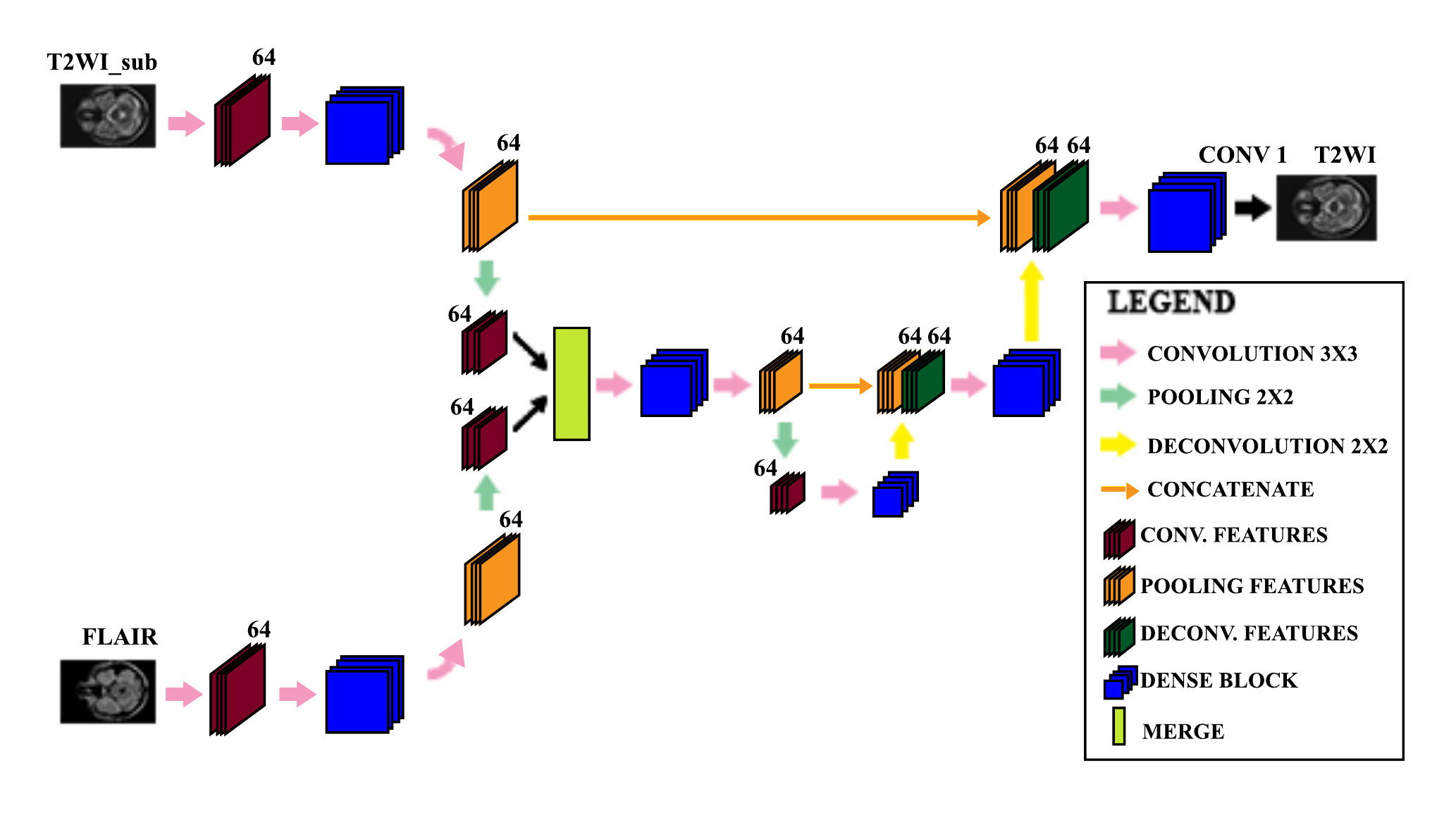}
	\caption{Scheme of the proposed Multimodal Dense U-Net architecture.}
	\label{fig:multidenseunet}
\end{figure}

Both inputs initially undergo separate contraction transformations to then merge later and follow the classic coding-decoding approach of U-Net models. The proposed Multimodal Dense U-Net is depicted in Fig.~\ref{fig:multidenseunet}.

The network consists essentially of 4 components, namely convolutive layers, pooling layers, deconvolutive layers and dense blocks. The size of the characteristic map decreases along the contraction path through the pooling blocks as it increases along the expansion path by deconvolution. Pooling partitions the input image into a set of squares, and for each of the resulting regions returns the maximum value as output. Its purpose is to progressively reduce the size of the representations, so as to reduce the number of parameters and the computational complexity of the network, at the same time counteracting any overfitting.

Deconvolutive layers act inversely with respect to pooling and aim to increase the spatial dimensions of the inputs. This allows to obtain images of a size comparable to those of the input images from the network. In the simplest case these levels can be implemented as static oversampling with bilinear interpolation.

The dense block, proposed in \cite{HuangCVPR2017}, allows to effectively increase the depth of the entire network while maintaining a low complexity. Moreover, it requires less parameters to be trained. The dense block consists of three consecutive operations: batch normalization (BN) 
, ELUs activation functions \cite{ClevertICLR2016} and $3 \times 3$ convolution filters. The hyper-parameters for the dense block are the growth rate (GR) and the number of convolutional layers (NC). The network ends with a reconstruction level consisting of a dense block followed by a $1 \times 1$ convolutional layer that yields the reconstructed T2WI.
%
%
%
%
%
%
\section{Experimental Results}
\label{sec:results}
\subsection{Dataset and Network Setting}
\label{subs:dataset}
We test the proposed network on a dataset containing MRIs of multiple sclerosis patients \cite{ZigaNI2018}. In particular, the dataset is related to 30 patients and it contains axial 2D-T1W, 2D-T2W and 3D-FLAIR images. 
The final voxel size of such images is $0.46 \times 0.46 \times 0.8$ mm$^{3}$. In our work, a further preprocessing has been performed in MATLAB to make the voxel size isotropic to $0.8 \times 0.8 \times 0.8$ mm$^{3}$, to extract slices of size $192 \times 292$, and to shrink intensity to the range $\left[0, 1\right]$. T2WIsub images were created by considering two types of masks, the center mask and the proposed custom mask, with a subsampling factor $k = 4$.

The proposed Multimodal Dense U-Net has been implemented on Keras. In the training stage, for each patient we provide the network with 150 FLAIR and T2WIsub images using the T2WIs as target. For dense blocks, we set a zero growth rate and a number of levels equal to 5 with feature maps size of 64 and ELU activation levels. We use Adam as an optimizer for training. A total of 80 epochs are performed, with early stopping. The duration of each epoch is about 15 minutes, having set a batch size of 4 and using a desktop PC with an Intel Core i5 6600-K 3.50 GHz CPU, 16 GB of RAM and NVIDIA GeForce GTX 970 GPU. To quantitatively evaluate reconstruction performance, we use the MSE and DSSIM metrics.
\begin{figure}[t]
	\centering
	\subfigure[]{\includegraphics[width=0.8\textwidth,keepaspectratio]{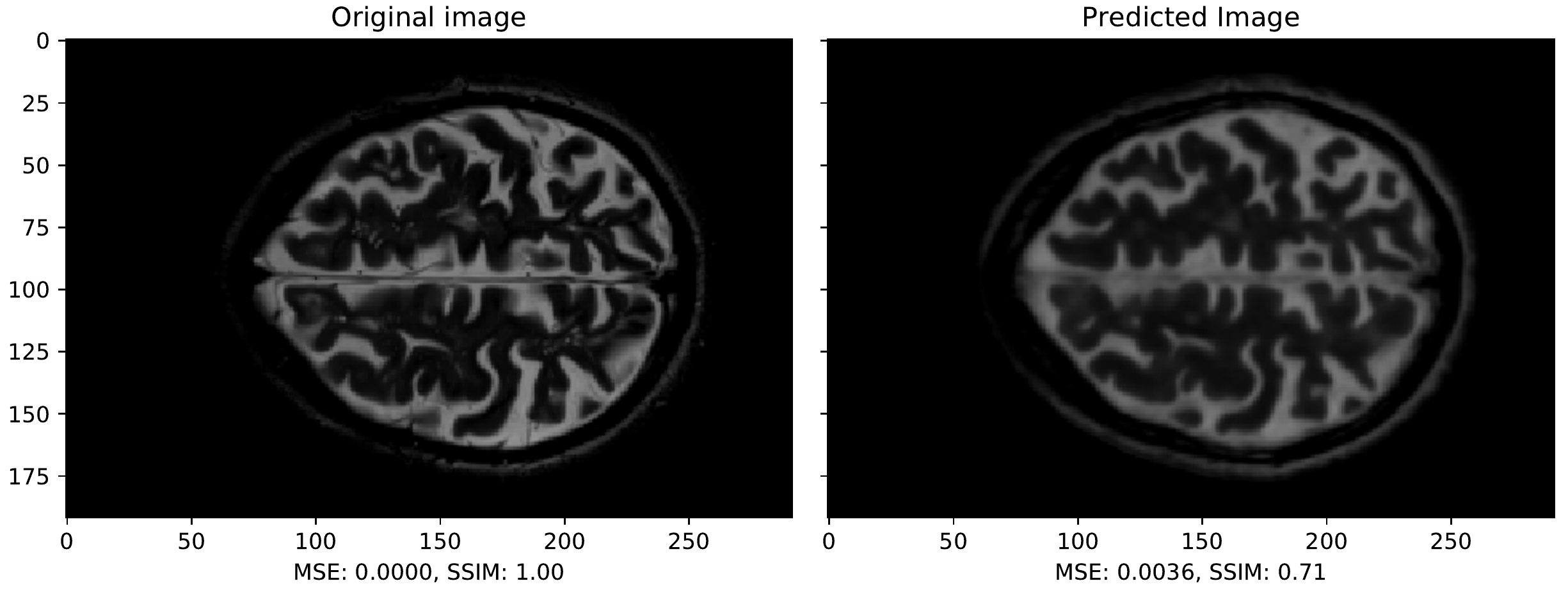}\label{fig:centermaskresults}}\\
\subfigure[]{\includegraphics[width=0.8\textwidth,keepaspectratio]{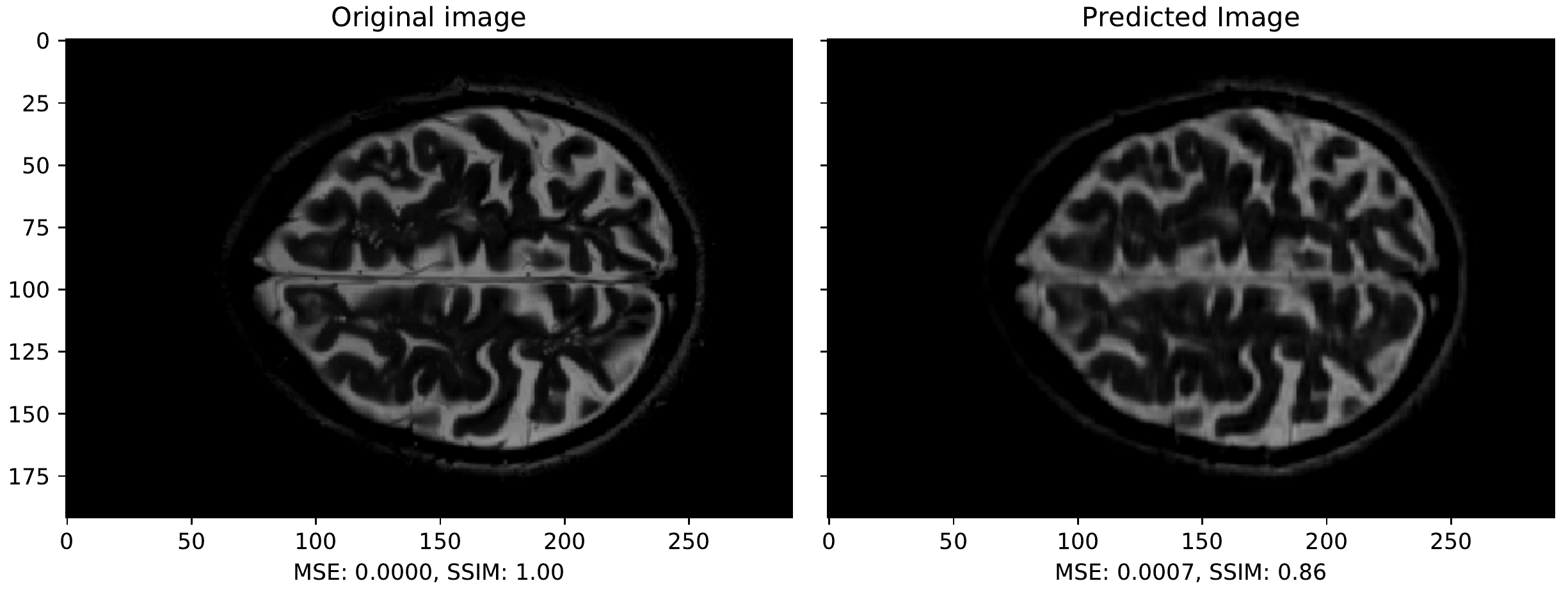}\label{fig:custommaskresults}}
	\caption{Predicted images using: a) center mask and b) the proposed custom mask.}
	\label{fig:maskresults}
\end{figure}
\subsection{Evaluation of the Proposed Mask}
\label{subs:maskevaluation}
We want to evaluate first the effectiveness of the proposed custom subsampling mask compared to the center mask on the quality of reconstruction in terms of the SSIM using a Dense U-Net network.

Results are shown in Fig.~\ref{fig:maskresults}, where it is clear that the proposed custom mask allows us to obtain a reconstruction of the image with outstanding performance. In particular, using the center mask we get a 71\% reconstruction percentage compared to the target (Fig.~\ref{fig:centermaskresults}), while using the proposed custom mask (Fig.~\ref{fig:custommaskresults}) the similarity index rises up to 86\%.
\subsection{Evaluation of the Proposed Deep Architecture}
\label{subs:networkevaluation}
Conceptually, the proposed architecture and the standard Dense U-Net might appear similar, but the former one manages the two inputs differently. 
Moreover, the hyper-parameters of the dense blocks chosen for our network considerably change the concept of dense block as the whole of growth was set to zero thus avoiding internal expansion in dense blocks.

\begin{figure}[t]
	\centering
	\includegraphics[width=0.8\textwidth,keepaspectratio]{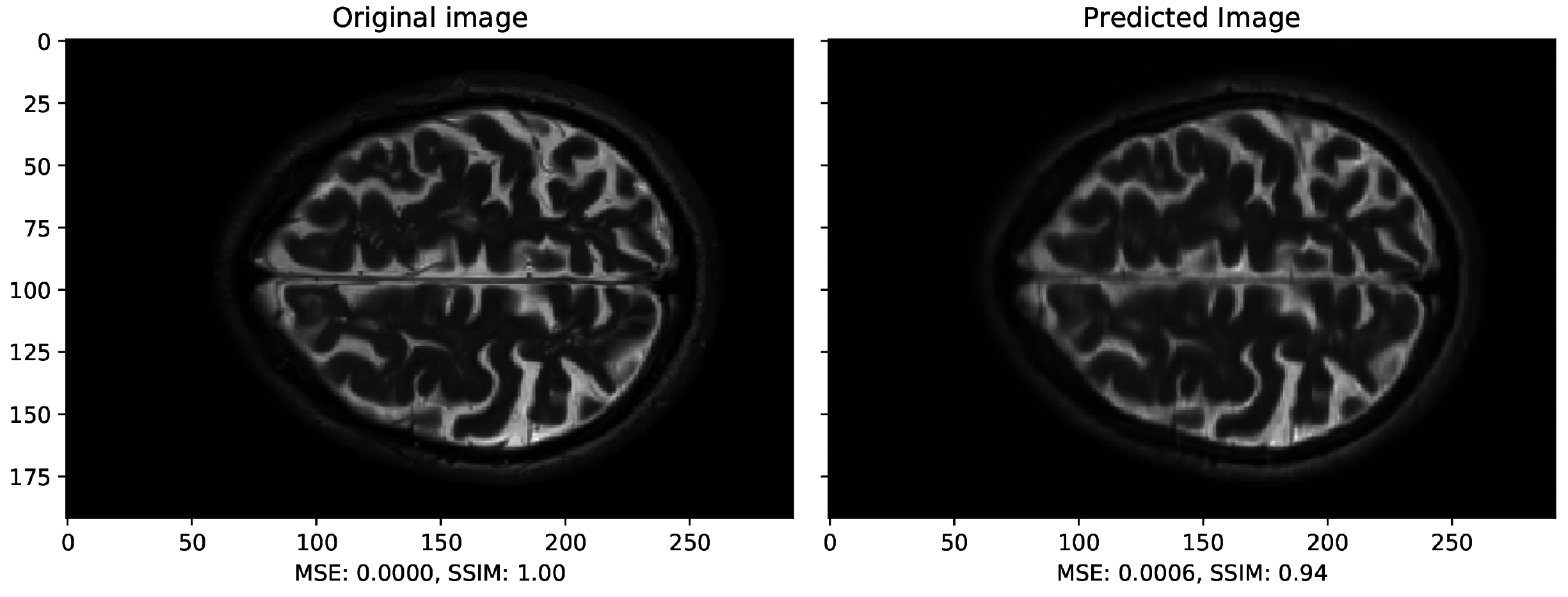}	 
	\caption{Predicted T2WI reconstructed by the proposed Multimodal Dense U-Net.}
	\label{fig:mduresults}
\end{figure}

\begin{figure}[t]
	\centering
	\includegraphics[width=0.75\textwidth,keepaspectratio]{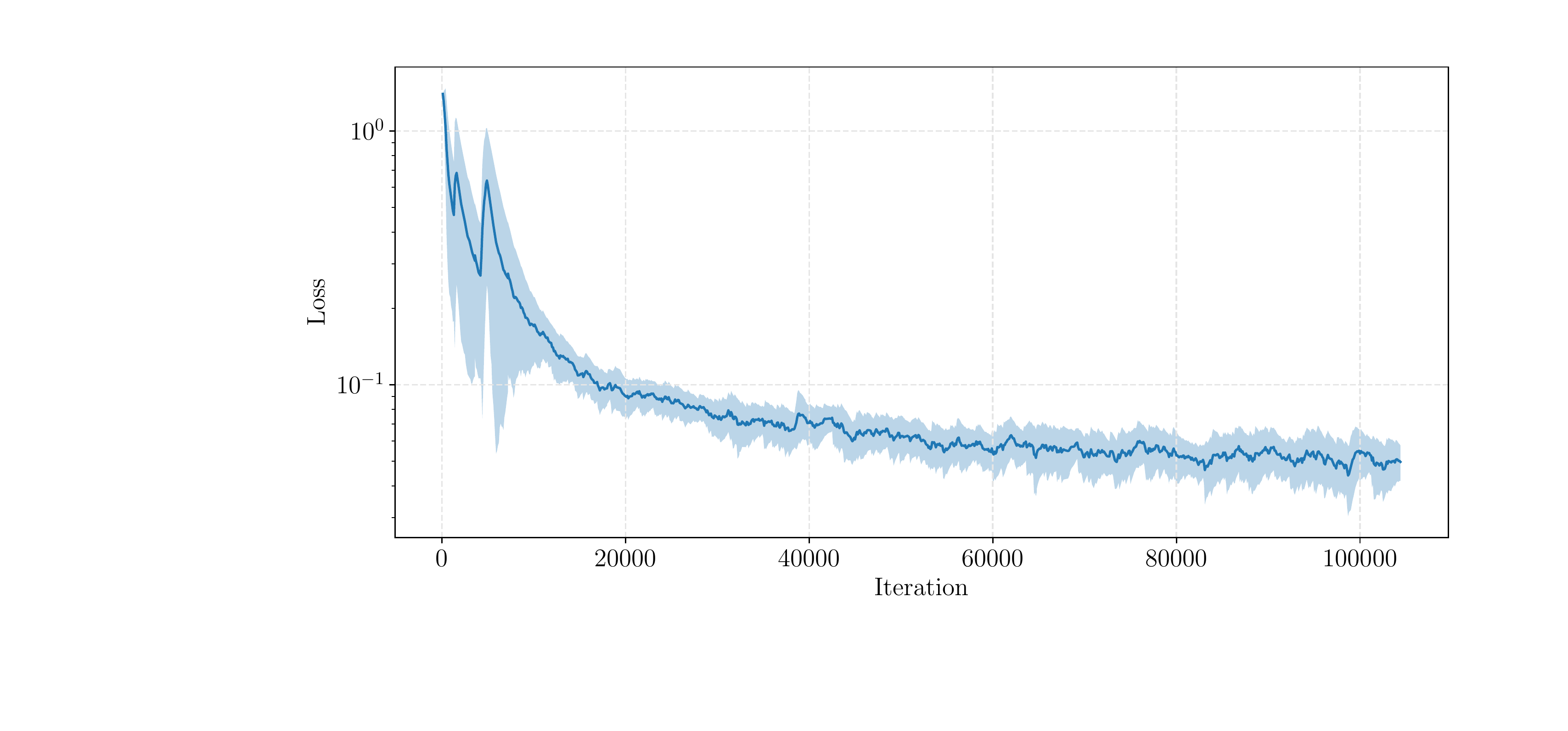}
	\caption{Loss function behavior.}
	\label{fig:loss}
\end{figure}

We compare the reconstruction quality of the two networks in terms of SSIM having used the mask that provided the best performance for the subsampling, i.e., the proposed custom mask. Results are shown in Fig.~\ref{fig:mduresults}, where it is clear that the quality of reconstruction has been considerably improved compared to Dense U-Net. In particular, the degree of similarity with respect to the target is 94\% rather than 86\% of the Dense U-Net. By using the proposed architecture, high image quality is achieved, thus enabling the recognition of brain injuries caused by the disease. We also show the loss function behavior for the proposed method in Fig.~\ref{fig:loss}.

%
%
%
\section{Conclusion}
\label{sec:concl}
In this work, we propose a deep learning model exploiting the capabilities of both multimodal networks and dense blocks. In particular, the proposed approach allows to reconstruct T2WIs, subsampled by a factor of 4, thus leveraging the correlation that exists with FLAIR images. At the same time, the proposed method is able to maintain a high quality of image reconstruction, in particular in the area of the brain lesions due to multiple sclerosis. The comparison with a state-of-the-art Dense U-Net architecture has shown that the proposed network outperforms both in terms of perceptive quality and in terms of execution times. Future works will focus on increasing the speed of the MRI scan, with the goal of achieving an acceleration of at least 10 times, and on further improving the reconstruction quality.
%

\begin{thebibliography}{10}
\providecommand{\url}[1]{\texttt{#1}}
\providecommand{\urlprefix}{URL }

\bibitem{Bell1984}
Beall, P.T., Amtey, S.R., Kasturi, S.R.: {NMR} Data Handbook for Biomedical
  Applications. Pergamon Books Inc., Elmsford, NY (1984)

\bibitem{ClevertICLR2016}
Clevert, D.A., Unterthiner, T., Hochreiter, S.: Fast and accurate deep network
  learning by exponential linear units ({ELU}s). In: International Conference
  on Learning Representations (ICLR). pp. 1--14. San Juan, Puerto Rico (May
  2016)

\bibitem{GamperMRM2008}
Gamper, U., Boesiger, P., Kozerke, S.: Compressed sensing in dynamic {MRI}.
  Magnetic Resonance in Medicine  59(2),  365--373 (Feb 2008)

\bibitem{Haacke1999}
Haacke, E.M., Brown, R.W., Thompson, M.R., Venkatesan, R.: Magnetic Resonance
  Imaging: {P}hysical Principles and Sequence Design, vol.~82. Wiley-Liss, New
  York, NY (1999)

\bibitem{HuangCVPR2017}
Huang, G., Liu, Z., van~der Maaten, L., Weinberger, K.Q.: Densely connected
  convolutional networks. In: IEEE Conference on Computer Vision and Pattern
  Recognition (CVPR). pp. 2261--2269. Honolulu, HI (Jul 2017)

\bibitem{HuangMRI2014}
Huang, J., Chen, C., Axel, L.: Fast multi-contrast {MRI} reconstruction.
  Magnetic Resonance Imaging  32(10),  1344--1352 (Dec 2014)

\bibitem{JaspanBJR2015}
Jaspan, O.N., Fleysher, R., Lipton, M.L.: Compressed sensing {MRI}: {A} review
  of the clinical literature. The British Journal of Radiology  88(1056) (Dec
  2015)

\bibitem{JinTIP2017}
Jin, K.H., McCann, M.T., Froustey, E., Unser, M.: Deep convolutional neural
  network for inverse problems in imaging. IEEE Transactions on Image
  Processing  26(9),  4509--4522 (Sep 2017)

\bibitem{KimMP2018}
Kim, K.H., Do, W.J., Park, S.H.: Improving resolution of {MR} images with an
  adversarial network incorporating images with different contrast. Medical
  Physics  45(7),  3120--3131 (Jul 2018)

\bibitem{Liang2000}
Liang, Z.P., Lauterbur, P.C.: Principles of Magnetic Resonance Imaging. A
  Signal Processing Perspective. The Institute of electrical and Electronics
  Engineers, New York, NY (2000)

\bibitem{LustigMRM2007}
Lustig, M., Donoho, D., Pauly, J.M.: Sparse {MRI}: {T}he application of
  compressed sensing for rapid {MR} imaging. Magnetic Resonance in Medicine
  58,  1182--1195 (Oct 2007)

\bibitem{LustigSPM2008}
Lustig, M., Donoho, D.L., Santos, J.M., Pauly, J.M.: Compressed sensing {MRI}.
  IEEE Signal Processing Magazine  25(2),  72--82 (Mar 2008)

\bibitem{MccannSPM2017}
McCann, M.T., Jin, K.H., Unser, M.: Convolutional neural networks for inverse
  problems in imaging: {A} review. IEEE Signal Processing Magazine  34(6),
  85--95 (Nov 2017)

\bibitem{QinTMI2019}
Qin, C., Schlemper, J., Caballero, J., Price, A.N., Hajnal, J.V., Rueckert, D.:
  Convolutional recurrent neural networks for dynamic {MR} image
  reconstruction. IEEE Transactions on Medical Imaging  38(1),  280--290 (Jan
  2019)

\bibitem{RonnenbergerMICCAI2015}
Ronnenberger, O., Fischer, P., Brox, T.: U-{N}et: {C}onvolutional networks for
  biomedical image segmentation. In: International Conference on Medical Image
  Computing and Computer-Assisted Intervention (MICCAI). Lecture Notes in
  Computer Science, vol. 9351, pp. 234--241. Springer, Cham (2015)

\bibitem{RoyARXIV2018}
Roy, S., Butman, J.A., Reich, D.S., Calabresi, P.A., Pham, D.L.: Multiple
  sclerosis lesion segmentation from brain {MRI} via fully convolutional neural
  networks. arXiv preprint arXiv:1803.09172v1  (Mar 2018)

\bibitem{SchlmperTMI2018}
Schlemper, J., Caballero, J., Hajnal, J.V., Price, A.N., Rueckert, D.: A deep
  cascade of convolutional neural networks for dynamic {MR} image
  reconstruction. IEEE Transactions on Medical Imaging  37(2),  491--503 (Feb
  2018)

\bibitem{ZigaNI2018}
\v{Z}iga, L., Galimzianova, A., Koren, A., Lukin, M., Pernu\v{s}, F., Likar,
  B., \v{S}piclin, v.: A novel public {MR} image dataset of multiple sclerosis
  patients with lesion segmentations based on multi-rater consensus.
  Neuroinformatics  16(1),  51--63 (Jan 2018)

\bibitem{XiangTBE2018}
Xiang, L., Chen, Y., Chang, W., Zhan, Y., Lin, W., Wang, Q., Shen, D.: Deep
  leaning based multi-modal fusion for fast {MR} reconstruction. IEEE
  Transactions on Biomedical Engineering  (Early Access) (2018)

\end{thebibliography}

%
\end{document}